\documentclass[aps,prd,onecolumn,groupedaddress,showpacs,nofootinbib,amssymb]{revtex4}
\usepackage[dvips]{graphicx}
\usepackage{amssymb}
\usepackage{amsmath}
\usepackage{graphicx,,color}
\usepackage{amsfonts}
\usepackage{bm}
\usepackage{cancel}
\usepackage{comment}

\newcommand\be{\begin{equation}}
\newcommand\ee{\end{equation}}

\allowdisplaybreaks[4]

\begin{document}

\title{Running of the Spectral Index and Inflationary Dynamics of $F(R)$ Gravity}
\author{S.D. Odintsov,$^{1,2}$\,\thanks{odintsov@ice.cat}
V.K. Oikonomou,$^{3}$\,\thanks{v.k.oikonomou1979@gmail.com}}
\affiliation{$^{1)}$ ICREA, Passeig Luis Companys, 23, 08010 Barcelona, Spain\\
$^{2)}$ Institute of Space Sciences (IEEC-CSIC) C. Can Magrans
s/n,
08193 Barcelona, Spain\\
$^{3)}$Department of Physics, Aristotle University of
Thessaloniki, Thessaloniki 54124, Greece}

\tolerance=5000

\begin{abstract}
In this work we shall provide a model-independent general
calculation of the running of the spectral index for vacuum $F(R)$
gravities. We shall exploit the functional form of the spectral
index and of the tensor-to-scalar ratio in order to present a
general $n_s-r$ relation for vacuum $F(R)$ gravity theories. As we
show, viable $F(R)$ gravity theories can be classified to two
classes of models regarding their prediction for the running
spectral index. The $R^2$-attractor models predict a running of
the spectral index in the range $-10^{-3}<a_s<-10^{-4}$, which
classifies them in the same universality class that most
inflationary scalar field models belong to. We provide three
models of this sort, for which we verify our claims in detail.
However there exist viable $F(R)$ gravity models with running of
spectral index outside the range $-10^{-3}<a_s<-10^{-4}$ and in
some cases it can be positive. We also present an $R^2$-corrected
scalar field model, which also predicts a running of the spectral
index in the range $-10^{-3}<a_s<-10^{-4}$. For all the cases we
studied, we found no evidence for the most phenomenologically
interesting scenario of having $r<10^{-4}$ and a running
$a_s<-10^{-3}$, which in principle could be realized.
\end{abstract}


\maketitle

\section{Introduction}

The next two decades are expected to be the most sensational for
theoretical cosmology, theoretical physics and theoretical
astrophysics. Many experiments are expected to finally commence,
which will probe the primordial Universe, and are expected to shed
light to mysterious hypothetical eras, such as the inflationary
era and the primordial dark ages that start from the reheating era
and beyond. There are a variety of experiments with the first
being the stage four Cosmic Microwave Background (CMB)
experiments, such as the CMB-S4 \cite{CMB-S4:2016ple} and the
Simons observatory \cite{SimonsObservatory:2019qwx}. These
experiments will probe directly the $B$-modes of inflation, or
will put tighter constraints on the inflationary parameters. Apart
from these two experiments, in the 2030's decade several
interferometers and other experiments
\cite{Hild:2010id,Baker:2019nia,Smith:2019wny,Crowder:2005nr,Smith:2016jqs,Seto:2001qf,Kawamura:2020pcg,Bull:2018lat}
will seek for a stochastic primordial gravitational wave
background.

It is apparent that inflation will be put into strong tests and
will be scrutinized to a great extent during the next two decades.
Inflation is an appealing hypothetical scenario for the primordial
era, which solves all the problems of the standard Big Bang
cosmology \cite{inflation1,inflation2,inflation3,inflation4}.
However it is possible that no primordial gravitational waves
signal is discovered nor a direct $B$-mode is detected by the
experiments. This will put inflation to a severely difficult
position, and it is highly likely for this to occur, if inflation
is controlled solely by a scalar field or some simple modified
gravity, such as $f(R)$ gravity. This is because both the
aforementioned theories yield an undetectable prediction for the
energy spectrum of the primordial gravitational waves, unless the
reheating era is abnormal, in which case a signal will be detected
for the aforementioned theories too. Of course, there exist a
large variety of modified gravity models that can accommodate the
observational data
\cite{reviews1,reviews1a,reviews2,reviews3,reviews4}, but in the
case of complete absence of future observations, things will be
quite difficult for cosmologists. In view of this pessimistic but
nevertheless possible scenario, it is vital to have a backup plan
looking for inflation in other parameters than the CMB
constraints. One parameter with profound importance is the running
of the spectral index $a_s$. Most of the inflationary single
scalar field models predict that $a_s$ is in the range
$-10^{-3}<a_s<-10^{-4}$. Thus if the running of the spectral index
is ever discovered in this range, it is highly likely that the
underlying inflationary theory might be a single scalar field
theory. The single scalar field descriptions of inflation form a
universality class in which many models belong. In fact, in some
cases it is possible to detect a signal for the running of the
spectral index in the next generation of CMB experiments, without
even detecting the $B$-modes of inflation directly (thus without
knowing the tensor-to-scalar ratio) \cite{Easther:2021eje}. This
scenario also corresponds to another universality class of single
scalar field models \cite{Easther:2021eje}.

In view of the importance of the running of the spectral index for
future experiments, in this article we shall calculate the most
general form of the running of the spectral index for vacuum
$F(R)$ gravity in the Jordan frame
\cite{Nojiri:2003ft,Capozziello:2005ku,Capozziello:2004vh,Capozziello:2018ddp,Hwang:2001pu,Cognola:2005de,Nojiri:2006gh,Song:2006ej,Capozziello:2008qc,Bean:2006up,Capozziello:2012ie,Faulkner:2006ub,Olmo:2006eh,Sawicki:2007tf,Faraoni:2007yn,Carloni:2007yv,
Nojiri:2007as,Capozziello:2007ms,Deruelle:2007pt,Appleby:2008tv,Dunsby:2010wg,Odintsov:2020nwm,Odintsov:2019mlf,Odintsov:2019evb,Oikonomou:2020oex,Oikonomou:2020qah}.
Using a model independent approach, which will enable us to
quantify in simple parameters the viability of the $F(R)$ gravity,
we shall calculate the running of the spectral index for a general
vacuum $F(R)$ gravity. The viability constraint of the vacuum
$F(R)$ gravity will constrain appropriately the possible values of
the running of the spectral index. As we shall see, there are two
classes of viable vacuum $F(R)$ gravity theories, one is the $R^2$
attractors and the other class deviates from this class
significantly. As we shall demonstrate, the $R^2$ attractors yield
a running of the spectral index in the range
$-10^{-3}<a_s<-10^{-4}$, thus vacuum $F(R)$ gravity theories also
belong to the universality class that the most inflationary scalar
field models belong too. We exemplify this important class of
models by using three characteristic $F(R)$ gravity models, and we
explicitly calculate the running of the spectral index for these
models to further support our model-independent approach. Also the
viable $F(R)$ gravities which deviate from the $R^2$-attractors do
not belong to the aforementioned universality class, and for these
theories the predictions for the running is larger than
$|a_s|>10^{-3}$ and in some cases it can be positive. Thus with
this work we aim to point out the fact that some classes of $F(R)$
gravity models are indistinguishable from single scalar field
models. However there exist viable $F(R)$ models that deviate from
the single scalar field inflationary universality class. We also
present in brief an $R^2$-corrected scalar field theory, and we
calculate the running of the spectral index in this class of
theories. $R^2$ corrections in scalar theories are frequently
studied in the literature
\cite{Ema:2020evi,Ivanov:2021ily,Gottlober:1993hp,delaCruz-Dombriz:2016bjj,Enckell:2018uic,Karam:2018mft,Kubo:2020fdd,Gorbunov:2018llf,Calmet:2016fsr,Oikonomou:2021msx},
and are highly motivated because the effective inflationary
Lagrangian described by a single scalar field might have
$R^2$-corrections. As we shall demonstrate, it is possible for
$R^2$-corrected scalar field models to belong to the universality
class of the inflationary single scalar field models.

This paper is organized as follows: In section II we present the
essential features of vacuum $F(R)$ gravity theory. We calculate
the spectral index of the primordial curvature perturbations and
the tensor-to-scalar ratio and we provide a model independent
universal $n_s-r$ relation for $F(R)$ gravity. A thorough study of
the running of the spectral index, along with explicit examples is
presented too. In section III we perform the same analysis for
$R^2$-corrected single scalar field theories and the conclusions
are presented in the last section of the article.

For the purposes of this article, we shall assume that the
geometric background is that of a flat Friedmann-Robertson-Walker
metric, with line element,
\begin{equation}
\label{JGRG14} ds^2 = - dt^2 + a(t)^2 \sum_{i=1,2,3}
\left(dx^i\right)^2\, ,
\end{equation}
where $a(t)$ is the scale factor.

\section{The Vacuum $F(R)$ Gravity Inflationary Dynamics and the Running of the Spectral Index}

As we mentioned in the introduction, one of our aims in this
article is to investigate whether $F(R)$ gravity produces a
running of the spectral index $a_s$ of the primordial scalar
perturbations in the range $-10^{-3}<a_s<-10^{-4}$. Many viable
scalar field theories produce a running of the spectral index
belonging in the above range, so these basically constitute a
characteristic universality class of models. As we will show in
this section, using a model independent approach, some classes of
$F(R)$ gravity models belong in the same universality class as the
scalar models, however, it is possible to have viable $F(R)$
gravity models which deviate from the universality class.

To start off, we consider a vacuum $F(R)$ gravity theory with its
gravitational action being,
\begin{equation}\label{action1dse}
\mathcal{S}=\frac{1}{2\kappa^2}\int \mathrm{d}^4x\sqrt{-g}F(R),
\end{equation}
where $\kappa^2=8\pi G=\frac{1}{M_p^2}$ and with $M_p$ denoting
the reduced Planck mass. In the context of the metric formalism,
upon varying the gravitational action with respect to the metric,
we obtain the following field equations,
\begin{equation}\label{eqnmotion}
F_R(R)R_{\mu \nu}(g)-\frac{1}{2}F(R)g_{\mu
\nu}-\nabla_{\mu}\nabla_{\nu}F_R(R)+g_{\mu \nu}\square F_R(R)=0\,
,
\end{equation}
with $F_R=\frac{\mathrm{d}F}{\mathrm{d}R}$. Eq. (\ref{eqnmotion})
can be rewritten as follows,
\begin{align}\label{modifiedeinsteineqns}
R_{\mu \nu}-\frac{1}{2}Rg_{\mu
\nu}=\frac{\kappa^2}{F_R(R)}\Big{(}T_{\mu
\nu}+\frac{1}{\kappa^2}\Big{(}\frac{F(R)-RF_R(R)}{2}g_{\mu
\nu}+\nabla_{\mu}\nabla_{\nu}F_R(R)-g_{\mu \nu}\square
F_R(R)\Big{)}\Big{)}\, .
\end{align}
Using the FRW metric (\ref{JGRG14}), the field equations take the
following form,
\begin{align}
\label{JGRG15} 0 =& -\frac{F(R)}{2} + 3\left(H^2 + \dot H\right)
F_R(R) - 18 \left( 4H^2 \dot H + H \ddot H\right) F_{RR}(R)\, ,\\
\label{Cr4b} 0 =& \frac{F(R)}{2} - \left(\dot H +
3H^2\right)F_R(R) + 6 \left( 8H^2 \dot H + 4 {\dot H}^2 + 6 H
\ddot H + \dddot H\right) F_{RR}(R) + 36\left( 4H\dot H + \ddot
H\right)^2 F_{RRR} \, ,
\end{align}
where $F_{RR}=\frac{\mathrm{d}^2F}{\mathrm{d}R^2}$, and
$F_{RRR}=\frac{\mathrm{d}^3F}{\mathrm{d}R^3}$. Moreover, $H$
denotes as usual the Hubble rate $H=\dot a/a$ and the Ricci scalar
for the FRW metric (\ref{JGRG14}) is $R=12H^2 + 6\dot H$.

In order to reveal the functional form of the running of the
spectral index for vacuum $F(R)$ theories in a model-independent
way, one needs to find the general form of the $n_s-r$ relation,
where $n_s$ and $r$ are the spectral index of the scalar
primordial perturbations and $r$ is the tensor-to-scalar ratio.
This analysis was performed in Ref. \cite{Odintsov:2020thl},
however for the sake of self-completeness, we present here in
brief the general $n_s-r$ relation for vacuum $F(R)$ gravity. We
shall assume that the slow-roll approximations hold true,
\begin{equation}\label{slowrollconditionshubble}
\ddot{H}\ll H\dot{H},\,\,\, \frac{\dot{H}}{H^2}\ll 1\, .
\end{equation}
For vacuum $F(R)$ gravity, the slow-roll indices are defined as
follows \cite{Hwang:2005hb,Kaiser:1994vs,reviews1},
\begin{equation}
\label{restofparametersfr}\epsilon_1=-\frac{\dot{H}}{H^2}, \quad
\epsilon_2=0\, ,\quad \epsilon_3= \frac{\dot{F}_R}{2HF_R}\, ,\quad
\epsilon_4=\frac{\ddot{F}_R}{H\dot{F}_R}\,
 .
\end{equation}
Accordingly, the spectral index and the tensor-to-scalar ratio for
a general $F(R)$ theory read
\cite{reviews1,Kaiser:1994vs,Hwang:2005hb},
\begin{equation}
\label{epsilonall} n_s=
1-\frac{4\epsilon_1-2\epsilon_3+2\epsilon_4}{1-\epsilon_1},\quad
r=48\frac{\epsilon_3^2}{(1+\epsilon_3)^2}\, .
\end{equation}
We can further simplify the tensor-to-scalar ratio for vacuum
$F(R)$ gravity by using the Raychaudhuri equation,
\begin{equation}\label{approx1}
\epsilon_1=-\epsilon_3(1-\epsilon_4)\, ,
\end{equation}
hence by using the slow-roll assumptions, we have
$\epsilon_1\simeq -\epsilon_3$, and in effect, the spectral index
becomes,
\begin{equation}
\label{spectralfinal} n_s\simeq 1-6\epsilon_1-2\epsilon_4\, ,
\end{equation}
and accordingly the tensor-to-scalar ratio becomes $r\simeq 48
\epsilon_3^2$, and in conjunction with the fact that
$\epsilon_1\simeq -\epsilon_3$, we finally have,
\begin{equation}
\label{tensorfinal} r\simeq 48\epsilon_1^2\, .
\end{equation}
The slow-roll index $\epsilon_4$ is the most complicated and the
most interesting one. We have,
\begin{equation}\label{epsilon41}
\epsilon_4=\frac{\ddot{F}_R}{H\dot{F}_R}=\frac{\frac{d}{d
t}\left(F_{RR}\dot{R}\right)}{HF_{RR}\dot{R}}=\frac{F_{RRR}\dot{R}^2+F_{RR}\frac{d
(\dot{R})}{d t}}{HF_{RR}\dot{R}}\, ,
\end{equation}
and due to the slow-roll assumptions, $\dot{R}$ is,
\begin{equation}\label{rdot}
\dot{R}=24\dot{H}H+6\ddot{H}\simeq 24H\dot{H}=-24H^3\epsilon_1\, .
\end{equation}
Combining Eqs. (\ref{rdot}) and (\ref{epsilon41}) we obtain,
\begin{equation}\label{epsilon4final}
\epsilon_4\simeq -\frac{24
F_{RRR}H^2}{F_{RR}}\epsilon_1-3\epsilon_1+\frac{\dot{\epsilon}_1}{H\epsilon_1}\,
,
\end{equation}
however $\dot{\epsilon}_1$ is,
\begin{equation}\label{epsilon1newfiles}
\dot{\epsilon}_1=-\frac{\ddot{H}H^2-2\dot{H}^2H}{H^4}=-\frac{\ddot{H}}{H^2}+\frac{2\dot{H}^2}{H^3}\simeq
2H \epsilon_1^2\, ,
\end{equation}
hence $\epsilon_4$ reads,
\begin{equation}\label{finalapproxepsilon4}
\epsilon_4\simeq -\frac{24
F_{RRR}H^2}{F_{RR}}\epsilon_1-\epsilon_1\, .
\end{equation}
Upon introducing the dimensionless parameter $x$, defined in the
following way,
\begin{equation}\label{parameterx}
x=\frac{48 F_{RRR}H^2}{F_{RR}}\, ,
\end{equation}
the parameter $\epsilon_4$ takes the final form,
\begin{equation}\label{epsilon4finalnew}
\epsilon_4\simeq -\frac{x}{2}\epsilon_1-\epsilon_1\, .
\end{equation}
Hence, upon substituting $\epsilon_4$ from Eq.
(\ref{epsilon4finalnew}) in Eq. (\ref{spectralfinal}), we have,
\begin{equation}\label{asxeto1}
n_s-1=-4\epsilon_1+x\epsilon_1\, ,
\end{equation}
and we can solve the above equation with respect to $\epsilon_1$
to obtain,
\begin{equation}\label{spectralasfunctionofepsilon1}
\epsilon_1=\frac{1-n_s}{4-x}\, .
\end{equation}
Upon combining Eqs. (\ref{spectralasfunctionofepsilon1}) and
(\ref{tensorfinal}), we obtain,
\begin{equation}\label{mainequation}
r\simeq \frac{48 (1-n_s)^2}{(4-x)^2}\, .
\end{equation}
The above relation is a universal relation holding true for all
the vacuum $F(R)$ gravity models. It is crucial to note that the
viability of an $F(R)$ gravity model depends on the values of the
parameter $x$ at the first horizon crossing of a mode during the
inflationary era, since $x$ affects both the values of $r$ via Eq.
(\ref{mainequation}) and of the spectral index via Eq.
(\ref{asxeto1}).

At this point, let us derive the running of the spectral index for
general vacuum $F(R)$ gravity. The running of the spectral index
is defined as,
\begin{equation}\label{runningdef}
a_s=\frac{\mathrm{d} n_s}{\mathrm{d} \ln k}\, ,
\end{equation}
where $k$ is the comoving wavenumber of a primordial mode. We can
rewrite $a_s$ as follows,
\begin{equation}\label{runningdef1}
a_s=\frac{\mathrm{d} n_s}{\mathrm{d} \ln k}=\frac{\mathrm{d}
n_s}{\mathrm{d}N}\frac{\mathrm{d}N}{\mathrm{d} \ln k}\, ,
\end{equation}
where $N$ is the $e$-foldings number, and by using
$\frac{\mathrm{d}N}{\mathrm{d} \ln k}=\frac{1}{1-\epsilon_1}$, the
final expression for the running of the spectral index $a_s$ is,
\begin{equation}\label{runningdefmainfinal}
a_s=\frac{1}{1-\epsilon_1}\frac{\mathrm{d} n_s}{\mathrm{d}N}\, .
\end{equation}
The above relation is general and holds for any $F(R,\phi)$ model,
but let us specify our analysis focusing on vacuum $F(R)$ gravity,
in which case we have,
\begin{equation}\label{dndns}
\frac{\mathrm{d} n_s}{\mathrm{d}N}=\frac{\mathrm{d}
\epsilon_1}{\mathrm{d}N}\left( -4+x\right)+\frac{\mathrm{d}
x}{\mathrm{d}N} \epsilon_1\, .
\end{equation}
The term $\frac{\mathrm{d} \epsilon_1}{\mathrm{d}N}$ for vacuum
$F(R)$ gravity is,
\begin{equation}\label{depsilon1}
\frac{\mathrm{d}
\epsilon_1}{\mathrm{d}N}=\frac{\dot{\epsilon}_1}{H}\, ,
\end{equation}
so by using (\ref{epsilon1newfiles}) we finally get,
\begin{equation}\label{finalepsilon1dot}
\frac{\mathrm{d} \epsilon_1}{\mathrm{d}N}=2\epsilon_1^2\, .
\end{equation}
Now let us calculate $\frac{\mathrm{d} x}{\mathrm{d}N}$ and after
some algebra we obtain,
\begin{equation}\label{finalform}
\frac{\mathrm{d} x}{\mathrm{d}N}=-1152
\frac{F_{RRRR}}{F_{RR}}H^4-2x\epsilon_1+\frac{x^2}{2}\epsilon_1\,
,
\end{equation}
where $F_{RRRR}=\frac{\partial^4 F}{\partial R^4}$. Thus the term
$\frac{\mathrm{d} n_s}{\mathrm{d}N}$ related to the running of the
spectral index for a general vacuum $F(R)$ gravity is equal to,
\begin{equation}\label{runningFRgeneral}
\frac{\mathrm{d} n_s}{\mathrm{d}N}=2\epsilon_1^2\left(x-4
\right)-1152
\frac{F_{RRRR}}{F_{RR}}H^4-2x\epsilon_1+\frac{x^2}{2}\epsilon_1 \,
.
\end{equation}
The above expression for the running of the spectral index has to
be evaluated at the first horizon crossing, and crucially depends
on the values of $x$ and on the values of
$\frac{F_{RRRR}}{F_{RR}}H^4$. Clearly the values of $x$ at horizon
crossing determine the viability of the $F(R)$ gravity model and
the $F(R)$ gravity models can be distinguished in attractors
depending the value of $x$. One important class of $F(R)$ gravity
models are the $R^2$ attractors, which can be obtained in the case
that $x=0$ or $x\ll 1$. The case $x=0$ corresponds to the $R^2$
gravity and for both the $x=0$ and $x\ll 1$ cases, the universal
relation (\ref{mainequation}) becomes $r\simeq 3(1-n_s)^2$. Also
the term $\frac{F_{RRRR}}{F_{RR}}H^4$ for the $R^2$ model, but
also when $x\ll 1$, the term $\frac{F_{RRRR}}{F_{RR}}H^4$ has also
to satisfy $\frac{F_{RRRR}}{F_{RR}}H^4\ll 1$ or to be zero in
order to have consistency with the condition $x\ll 1$. In effect,
all the $R^2$ attractor models ($x=0$ or $x\ll 1$) yield a running
of the spectral index which has the following approximate form at
leading order,
\begin{equation}\label{r2attractorsrunningofspectral}
a_s\simeq -\frac{8\epsilon_1^2}{1-\epsilon_1}\, .
\end{equation}
Since the $R^2$ attractor $F(R)$ gravity models are viable, this
means that their first slow-roll index $\epsilon_1$ satisfies the
Planck 2018 constraints \cite{Planck:2018jri} which constrains
$\epsilon_1<0.0097$. Hence for the $R^2$ attractor models, this
means that the running of the spectral index is of the order
$a_s\sim -7.44\times 10^{-4}$. The current Planck constraints on
the scale dependence of the spectral index are $a_s=-0.006\pm
0.0013$ thus all the $R^2$ attractor models are viable as
expected. More importantly, the $R^2$ attractors belong to the
same universality class that many scalar models belong to, which
predict $-10^{-3}<a_s<-10^{-4}$
\cite{Easther:2021eje,Adshead:2010mc}. Let us consider in brief
three characteristic examples of this category, with the first
being the pure $R^2$ model
\cite{Starobinsky:1980te,Bezrukov:2007ep}, and the second being a
deformation of the pure $R^2$ model,
\begin{equation}\label{purefrsquaredarkenerhgy}
f(R)=R+\frac{R^2}{36 M^2}-\gamma \left(
\frac{R}{m_s}\right)^{\alpha}\, ,
\end{equation}
with $0<\alpha<1$ and $\gamma \sim \mathcal{O}(\Lambda)$ where
$\Lambda$ is the present day cosmological constant. The model
(\ref{purefrsquaredarkenerhgy}) is known to produce a viable
$R^2$-like inflation and a viable dark energy era, see Ref.
\cite{Odintsov:2021wjz}. Finally we shall consider another
deformation of the $R^2$ model, namely,
\begin{equation}\label{frfinal}
f(R)=R+\frac{R^2}{6 M^2}+\frac{M^2}{6}-\frac{\mathcal{C}_2
\left(M^4+6 M^2 R+R^2\right) \left(\sqrt[4]{e} \sqrt{\pi }
\mathrm{Erf}\left(\frac{\sqrt{M^2+R}}{2 M}\right)+\frac{2 M
e^{-\frac{R}{4 M^2}} \left(3 M^2+R\right) \sqrt{M^2+R}}{M^4+6 M^2
R+R^2}\right)}{32 M^9}\, ,
\end{equation}
with the constant of integration $\mathcal{C}_2$ having mass
dimensions $[\mathcal{C}_2]=[m]^7$. The model (\ref{frfinal}) was
recently studied in Ref. \cite{Odintsov:2021wjz} and it was shown
that it can generate a viable $R^2$-like inflationary era. The
pure $R^2$ model yields $x=0$ but the other two deformations of
the $R^2$ model, namely models (\ref{purefrsquaredarkenerhgy}) and
(\ref{frfinal}) yield both $x\neq 0$ and
$\frac{F_{RRRR}}{F_{RR}}H^4\neq 0$. Let us now present in brief
the values of the running of the spectral index for the three
models, and for $N\sim 60$ $e$-foldings. We start of with the pure
$R^2$ model which has $x=0$ and $a_s\sim -0.0055$, and regarding
the model (\ref{frfinal}), without getting into much details, it
yields the same $a_s$ as the $R^2$ model with $x\sim
\mathcal{O}(10^{-127})$ and $\frac{F_{RRRR}}{F_{RR}}H^4\sim
10^{-60}$ for $H\sim 10^{16}$GeV at first horizon crossing.
Finally, the model (\ref{purefrsquaredarkenerhgy}) yields $a_s\sim
-0.0007035$ with $x\sim \mathcal{O}(10^{-29})$ and
$\frac{F_{RRRR}}{F_{RR}}H^4\sim 10^{-28}$. Thus all the models
indeed fall in the category of the $R^2$ attractors and predict
the same running for the spectral index as the universality
classed for the scalar models do, in the range
$-10^{-3}<a_s<-10^{-4}$.

Now let us proceed to the cases that $x$ is not small. In this
case there are three distinct possible scenarios, firstly $0<x\leq
1$, secondly $x\sim \mathcal{O}(4)$ and thirdly $x>1$ and $x\neq
4$ or larger, with $x$ being evaluated at the first horizon
crossing for $N\sim 60$. The last two scenarios do not correspond
to viable $F(R)$ gravities because the former leads to a blow-up
in the tensor-to-scalar ratio, while the latter although yielding
a small value for the tensor-to-scalar ratio, it yields a large
spectral index $n_s$. Hence the only case that is worth studying
is $x\sim \mathcal{O}(1)$. In this case, the model can be viable
and yield $n_s$ and $r$ compatible with the Planck constraints.
Indeed, the presence of the term $x\epsilon_1$ in the spectral
index (\ref{spectralasfunctionofepsilon1}), can affect the
spectral index even if $0<x\leq 1$. On the contrary, the running
of the spectral index contains the term $\sim x^2\epsilon_1$,
hence if $\epsilon_1=0.0097$ which is the largest value allowed
from the Planck 2018 data \cite{Planck:2018jri}, with $0<x\leq 1$,
the term $x^2\epsilon_1$ will be of the order $x^2\epsilon_1 \in
\,\,(0,0.006208)$ (values larger than $0.007$ in total for the
running of the spectral index are excluded by the Planck 2018
data). Thus the running of the spectral index is dramatically
affected in this case, since the running can be positive and
large, and certainly this class of $F(R)$ gravity models do not
fall into the same universality class that the most of the viable
inflationary scalar models and the $R^2$ attractors belong to.
However, although this class of models certainly exists
theoretically, we have no model that belongs to this class to
present, due to the lack of analyticity. Only a few $F(R)$ gravity
models can be worked out analytically, unless the Hubble rate is
given. The same conclusions can be obtained if $-1\geq x<0$, thus
this case too is covered by the conclusions obtained for the case
$0<x\leq 1$. Also notice that the case $-1\geq x<0$ yields the
same running for the spectral index as the $0<x\leq 1$, with
different spectral index though.

Thus in this section we basically demonstrated that there exist
two classes of viable $F(R)$ gravity models that can be classified
according to their predictions for the running of the spectral
index. The first class contains all $R^2$ attractor models with
$x=0$ or $x\ll 1$, while the second class contains models with
$0<x\leq 1$. The attractor models belong to the universality class
that most of the known inflationary scalar models belong to, while
the other viable $F(R)$ models constitute a class of their own,
with possible significant effects on the running of the spectral
index. None however belongs to the class presented in Ref.
\cite{Easther:2021eje}, where for a small $r$ with $r\leq
10^{-4}$, then $a_s<-10^{-3}$. If the small $r$ and $a_s<-10^{-3}$
combination is verified experimentally, and specifically if the
$a_s<-10^{-3}$ case is verified experimentally, this can be in
favor of inflation and of specific inflationary theories which
have too small tensor-to-scalar ratio to be detected. In the next
section we shall also present an $f(R,\phi)$ gravity which has a
large tensor-to-scalar ratio and also satisfies
$-10^{-3}<a_s<-10^{-4}$. As we show, that model too belongs to the
universality class of the scalar field inflationary models, and
also deviates from the class of models with $r\leq 10^{-4}$, and
$a_s<-10^{-3}$. This further supports the claim of
\cite{Easther:2021eje} that if a measurement occurs with
$a_s<-10^{-3}$ and no hint for tensor perturbations is found, then
this will indicate that a small $r$ scalar theory might be
controlling inflation.

\section{The Running of the Spectral Index in $f(R,\phi)$ Gravity: $R^2$ Quantum-corrected Scalar Field Theory}

In this section we shall consider an $f(R,\phi)$ gravity instead
of an vacuum $F(R)$ gravity which we considered in the previous
section. Specifically, this $f(R,\phi)$ gravity will be an $R^2$
quantum corrected scalar field theory. We shall be interested in
investigating whether this class of theories also belongs to the
universality classes that most of the scalar inflationary theories
belong, regarding their predictions on the running of the spectral
index. The full analysis of the $R^2$ quantum corrected scalar
field theory is developed elsewhere \cite{myref}, here we are
mainly interested on the running of the spectral index.

Scalar field theories are of fundamental importance since scalar
fields are natural predictions of string theory, which is to date
the most appealing UV completion of classical gravitation and
particle physics. Although string theory is rather impossible to
be verified experimentally on terrestrial accelerators, it might
be possible to find its imprints at the effective inflationary
Lagrangian. Consider the most general scalar Lagrangian,
\begin{equation}\label{generalscalarfieldaction}
\mathcal{S}_{\varphi}=\int
\mathrm{d}^4x\sqrt{-g}\left(\frac{1}{2}Z(\varphi)g^{\mu
\nu}\partial_{\mu}\varphi
\partial_{\nu}\varphi+\mathcal{V}(\varphi)+h(\varphi)R
\right)\, .
\end{equation}
The vacuum configuration of scalar fields compels the scalar
fields to be either minimally or conformally coupled
\cite{Codello:2015mba}. The quantum corrections of a minimally
coupled or of a conformally coupled scalar field Lagrangian can be
of the form \cite{Codello:2015mba},
\begin{align}\label{quantumaction}
&\mathcal{S}_{eff}=\int
\mathrm{d}^4x\sqrt{-g}\Big{(}\Lambda_1+\Lambda_2
R+\Lambda_3R^2+\Lambda_4 R_{\mu \nu}R^{\mu \nu}+\Lambda_5 R_{\mu
\nu \alpha \beta}R^{\mu \nu \alpha \beta}+\Lambda_6 \square R\\
\notag & +\Lambda_7R\square R+\Lambda_8 R_{\mu \nu}\square R^{\mu
\nu}+\Lambda_9R^3+\mathcal{O}(\partial^8)+...\Big{)}\, ,
\end{align}
which contains up to fourth order derivative and is diffeomorphic
invariant. We shall consider the $R^2$ corrections in this paper,
so the effective Lagrangian is of the form,
\begin{equation}
\label{action} \centering
\mathcal{S}=\int{d^4x\sqrt{-g}\left(\frac{
f(R)}{2\kappa^2}-\frac{1}{2}g^{\mu \nu}\partial_{\mu}\varphi
\partial_{\nu}\varphi-\mathcal{V}(\varphi)\right)}\, ,
\end{equation}
where,
\begin{equation}\label{fr}
    f(R)=R+\frac{R^2}{36M^2}\, ,
\end{equation}
where $M$ is a mass scale which in the context of $R^2$-corrected
scalar field theory is a free parameter, in contrast to the vacuum
$R^2$ model \cite{Starobinsky:1980te,Bezrukov:2007ep}. For the FRW
metric, the field equations read,
\begin{equation} \label{Friedmann}
    3 f_{R} \mathcal{H}^2=\frac{R f_{R} -f}{2}-3\mathcal{H} \dot F_{R}+\kappa^2\big(\frac{1}{2}\dot\varphi^2+\mathcal{V}(\varphi)\big) \ ,
\end{equation}
\begin{equation} \label{Raychad}
    -2 f_{R} \dot{\mathcal{H}} = \kappa^2 \dot\varphi^2 + \ddot f_{R} -\mathcal{H} \dot f_{R} \ ,
\end{equation}
\begin{equation} \label{fieldeqmotion}
   \ddot\varphi+3\mathcal{H}\dot\varphi+\mathcal{V}'=0 \ ,
\end{equation}
with the ``dot'' and the ``prime'' denoting differentiation with
respect to the cosmic time and the scalar field respectively,
while $f_{R}=\frac{\partial f}{\partial R}$. By assuming the
slow-roll conditions,
\begin{equation}\label{slowrollH}
    \dot{\mathcal{H}} \ll \mathcal{H}^2 \ , \ \ddot{\mathcal{H}} \ll \mathcal{H} \dot{\mathcal{H}},
\end{equation}
and also the following approximations,
\begin{equation}\label{approxH}
    \frac{\dot{\mathcal{H}}^2}{M^2} \ll \mathcal{H}^2,\,\,\,\frac{\dot{\mathcal{H}}^2}{M^2} \ll
    \mathcal{V}(\varphi)\, ,
\end{equation}
the field equations at leading order read (the details are given
in \cite{myref}):
\begin{equation} \label{Friedman}
    \mathcal{H}^2 \simeq \frac{\kappa^2 \mathcal{V}(\varphi)}{3}+\mathcal{O}(\frac{\kappa^2 \dot{\varphi}^2}{2}\mathcal{H}^2),
\end{equation}
\begin{equation} \label{Raycha}
     \dot{\mathcal{H}} \simeq -\frac{\kappa^2 \dot{\varphi}^2}{2}  -\frac{\kappa^4
     \dot{\varphi}^4}{4M^2}\, .
\end{equation}
Furthermore the approximate forms of the slow-roll indices for the
$R^2$-corrected scalar theory are \cite{myref},
\begin{equation}\label{e1}
    \epsilon_1=\frac{1}{2\kappa^2} \left( \left(\frac{\mathcal{V}'}{V}\right)^2 +\frac{1}{6M^2}{\left(\frac{\mathcal{V}'}{V}\right)}^2\frac{\mathcal{V}'^2}{V}\right)\, .
\end{equation}
\begin{equation}\label{e2}
    \epsilon_2=-\frac{\mathcal{V}''}{\kappa^2 V}+\epsilon_1 \, ,
\end{equation}
\begin{equation}\label{e3}
    \epsilon_3=\frac{\epsilon_1}{-1-\frac{3M^2}{2\mathcal{H}^2}+\frac{\epsilon_1}{2}} \,
    ,
\end{equation}
while $\epsilon_4$ is $\epsilon_4=\frac{\dot E}{2\mathcal{H}E}$,
with,
\begin{equation}\label{E}
    E=1+\frac{2R}{36M^2}+\frac{8}{3\kappa^2 M^4}\frac{\mathcal{H}^2 \dot{\mathcal{H}}^2}{\dot \varphi^2} ,
\end{equation}
\begin{equation}\label{dotE}
    \dot E=\frac{4 \mathcal{H} \dot{\mathcal{H}}}{3M^2} + \frac{16}{3 \kappa^2 M^4 \dot \varphi^4}\big( \mathcal{H} \dot{\mathcal{H}}^3 \dot \varphi^2-\mathcal{H}^2 \dot{\mathcal{H}}^2 \dot \varphi \ddot \varphi \big),
\end{equation}
and we omit the final form of $\epsilon_4$ for brevity. Having the
slow-roll indices, one ca easily confront the theory with the
Planck observations. The spectral index for the $f(R,\phi)$ theory
is the same as in the vacuum $F(R)$ gravity of the previous
section, however, the tensor-to-scalar ratio is different at
leading order in the slow-roll indices
\cite{Hwang:2005hb,reviews1},
\begin{equation}\label{r}
    r=16(\epsilon_1 + \epsilon_3)\, .
\end{equation}
Let us choose a simple potential which can be shown \cite{myref}
that it yields a viable phenomenology,
\begin{equation}\label{V}
    \mathcal{V}(\varphi)=\frac{\mathcal{V}_0}{\kappa^4}(\kappa \varphi)^2,
\end{equation}
with $\mathcal{V}_0$ being a dimensionless parameter. We set
$M=\beta/\kappa$ where $\beta$ is dimensionless. This theory is a
viable one, and the viability is guaranteed when $\mathcal{V}_0
\sim \mathcal{O}(10^{-13})$ and $\beta \sim \mathcal{O}(10^{-6})$.
Full details on the inflationary phenomenology of this model is
given in \cite{myref} and one set of parameters that yield
viability for this model is, $\mathcal{V}_0=9.37 \times 10^{-13}$,
$\beta=6.8 \times 10^{-6}$ and $N=60$, for which we get,
\begin{equation}\label{nsrexample}
   n_{\mathcal{S}}=0.96611,\,\,\,
   r=0.063968,\, .
\end{equation}
For the same set of values for the free parameters we obtain
$a_s=-0.00056$, which belongs in the range
$-10^{-3}<a_s<-10^{-3}$. This model, as all the viable $F(R)$
gravity models, belong to the same universality class of
inflationary scalar potentials. Hence, this result further
supports the argument of Ref. \cite{Easther:2021eje}, which states
that if a measurement of the running of the spectral index is
found with $a_s<-10^{-3}$ and no sign of tensor perturbations is
found, this will indicate that probably a scalar inflationary
theory controls the dynamics of inflation, with characteristics
different than the scalar models which constitute the universality
class. Such a future observation will shed further light on the
$a_s-r$ relation.

\section{Conclusions}

In this paper we provided a model independent theoretical
framework for vacuum $F(R)$ gravity in order to predict the most
general form of the running of the spectral index for these
theories. Exploiting the functional form of the spectral index and
of the tensor-to-scalar ratio for vacuum $F(R)$ gravity, we
presented a quite general $n_s-r$ relation holding true for all
vacuum $F(R)$ gravities. The viable $F(R)$ gravities can be
classified in two distinct classes of models, which are
characterized by small or $\mathcal{O}(1)$ values for the
parameter $x\sim \frac{F_{RRR}}{F_{RR}}H^2$ when it is evaluated
at the first horizon crossing during inflation. The small $x$
values belong to the $R^2$-attractor models, which have a running
spectral index which takes values $-10^{-3}<x<-10^{-4}$. This
feature classifies the $R^2$-attractor models in the same
universality class that most inflationary single scalar field
models belong to. We presented three models, which belong to this
class, all of which were deformations of the $R^2$ model. We
calculated in detail the running of the spectral index and we
verified that the result was compatible with the model independent
approach we used for this class of models. In contrast, the viable
$F(R)$ gravity models with $x\sim \mathcal{O}(1)$ predict a
running of the spectral index which does not belong to the range
$-10^{-3}<a_s<-10^{-4}$ and it is highly likely that it is also
positive. However, we did not provide any example for this class
of $F(R)$ gravity models, which must exist theoretically, however
the lack of analyticity prevented us from finding an example of
this sort. Finally, we also provided another theory which belongs
to the universality class of the inflationary scalar models,
namely an $R^2$-corrected single scalar field theory. In this case
we used an explicit example for the calculation of the running of
the spectral index, due to the perplexity of the theoretical
framework. As we demonstrated, this theory too predicts
$-10^{-3}<a_s<-10^{-4}$. Notably, in all the cases we studied, we
found no evidence for models which predict $r<10^{-4}$ and
$a_s<-10^{-3}$, a class of theories pointed out in Ref.
\cite{Easther:2021eje}. This would be a particularly interesting
scenario, but it seems that it is not easy to find analytically a
vacuum $F(R)$ gravity which can realize such a scenario. It is
thus worth investigating whether this scenario can be realized in
non-minimally coupled theories or in mimetic scalar field theories
or even in generalized $F(R,\phi)$ theories. We hope to address
this issue in the near future.

\section*{Acknowledgments}

This work was supported by MINECO (Spain), project
PID2019-104397GB-I00 (S.D.O).

\end{document}